\documentclass[aps,showpacs,prd,twocolumn]{revtex4-1}
\usepackage{amsfonts}
\usepackage{amssymb}
\usepackage{amsmath}
\usepackage{color}
\usepackage[usenames,dvipsnames]{xcolor}
\usepackage{float}
\usepackage{accents}
\usepackage{graphicx}
\usepackage{epstopdf}
\usepackage{ulem}
\usepackage[colorlinks=true,pdfstartview=FitV,linkcolor=blue,citecolor=blue,urlcolor=blue,breaklinks=true]{hyperref}

\begin{document}

\title{First-order solitons with internal structures in an extended Maxwell-$%
CP(2)$ model }
\author{J. Andrade$^{1}$, R. Casana$^{1}$, E. da Hora$^{2}$ and C. dos Santos%
$^{3}$.}

\affiliation{$^{1}${Departamento de F\'{\i}sica, Universidade Federal do Maranh\~{a}o,}\\
65080-805, S\~{a}o Lu\'{\i}s, Maranh\~{a}o, Brazil.\\
$^{2}$Coordenadoria Interdisciplinar de Ci\^{e}ncia e Tecnologia,\\
Universidade Federal do Maranh\~{a}o, {65080-805}, S\~{a}o Lu\'{\i}s, Maranh\~{a}o, Brazil{.}\\
$^{3}$Centro de F\'{\i}sica e Departamento de F\'{\i}sica e Astronomia,
Faculdade de Ci\^{e}ncias da Universidade do Porto, 4169-007, Porto,
Portugal.}

\begin{abstract}
We study a Maxwell-$CP(2)$ model coupled to a real scalar field through a
dielectric function multiplying the Maxwell term. In such a context, we look
for first-order rotationally symmetric solitons by means of the Bogomol'nyi
algorithm, i.e. by minimizing the total energy of the effective model. We
perform our investigation by choosing an explicit form of the dielectric
function. The numerical solutions show regular vortices whose shapes
dramatically differ from their canonical counterparts. We can understood
such differences as characterizing the existence of an internal structure.
\end{abstract}

\maketitle

\affiliation{$^{1}${Departamento de F\'{\i}sica, Universidade Federal do Maranh\~{a}o,}\\
65080-805, S\~{a}o Lu\'{\i}s, Maranh\~{a}o, Brazil.\\
$^{2}$Coordenadoria Interdisciplinar de Ci\^{e}ncia e Tecnologia,\\
Universidade Federal do Maranh\~{a}o, {65080-805}, S\~{a}o Lu\'{\i}s, Maranh\~{a}o, Brazil{.}\\
$^{3}$Centro de F\'{\i}sica e Departamento de F\'{\i}sica e Astronomia,
Faculdade de Ci\^{e}ncias da Universidade do Porto, 4169-007, Porto,
Portugal.}

\affiliation{$^{1}${Departamento de F\'{\i}sica, Universidade Federal do Maranh\~{a}o,}\\
65080-805, S\~{a}o Lu\'{\i}s, Maranh\~{a}o, Brazil.\\
$^{2}$Coordenadoria Interdisciplinar de Ci\^{e}ncia e Tecnologia,\\
Universidade Federal do Maranh\~{a}o, {65080-805}, S\~{a}o Lu\'{\i}s, Maranh\~{a}o, Brazil{.}\\
$^{3}$Centro de F\'{\i}sica e Departamento de F\'{\i}sica e Astronomia,
Faculdade de Ci\^{e}ncias da Universidade do Porto, 4169-007, Porto,
Portugal.}

\affiliation{$^{1}${Departamento de F\'{\i}sica, Universidade Federal do Maranh\~{a}o,}\\
65080-805, S\~{a}o Lu\'{\i}s, Maranh\~{a}o, Brazil.\\
$^{2}$Coordenadoria Interdisciplinar de Ci\^{e}ncia e Tecnologia,\\
Universidade Federal do Maranh\~{a}o, {65080-805}, S\~{a}o Lu\'{\i}s, Maranh\~{a}o, Brazil{.}\\
$^{3}$Centro de F\'{\i}sica e Departamento de F\'{\i}sica e Astronomia,
Faculdade de Ci\^{e}ncias da Universidade do Porto, 4169-007, Porto,
Portugal.} 
\affiliation{$^{1}${Departamento de F\'{\i}sica, Universidade Federal do Maranh\~{a}o,}\\
65080-805, S\~{a}o Lu\'{\i}s, Maranh\~{a}o, Brazil.\\
$^{2}$Coordenadoria Interdisciplinar de Ci\^{e}ncia e Tecnologia,\\
Universidade Federal do Maranh\~{a}o, {65080-805}, S\~{a}o Lu\'{\i}s, Maranh\~{a}o, Brazil{.}\\
$^{3}$Centro de F\'{\i}sica e Departamento de F\'{\i}sica e Astronomia,
Faculdade de Ci\^{e}ncias da Universidade do Porto, 4169-007, Porto,
Portugal.}

\section{Introduction}

\label{Intro}

In the context of classical models, topological solitons are commonly
described as time-independent solutions inherent to highly nonlinear field
theories \cite{n5}. For example, in (2+1)-dimensional gauge theories,
vortices stand for the rotationally symmetric solutions arising from the
Euler-Lagrange equations \cite{n1}. In addition, under special
circumstances, topological vortices can also be obtained via a particular
set of first-order differential equations (instead of the second-order
Euler-Lagrange ones). In general, such a first-order framework can be found
via a minimization procedure of the total energy of the system, which allows
to get a well-defined lower bound for the energy itself \cite{n41}-\cite{n4}%
. In this sense, first-order vortices were already verified to occur not
only in the usual Maxwell-Higgs scenario \cite{n4}, but also in the
Chern-Simons \cite{cshv} and in the composite Maxwell-Chern-Simons-Higgs one 
\cite{n42}.

The study of the first-order solitons arising from a $CP(N-1)$ model is
particularly important due to its straight phenomenological connection with
the Yang-Mills model defined in four dimensions, as explained in \cite{cpn-1}%
. Under such a perspective, the existence of first-order vortices in a
gauged $CP(2)$ scenario endowed with the usual Maxwell's action was firstly
suggested in \cite{loginov}, being explicitly demonstrated in \cite{casana}.
Moreover, electrically charged first-order vortices were also verified to
occur in a gauged $CP(2)$ model in the presence of both the Chern-Simons 
\cite{vini} and the composite Maxwell-Chern-Simons actions \cite{neyver},
separately.

At the same time, in the context of extended field models, it is interesting
to point out that the standard Maxwell-Higgs one was itself enlarged in
order to accommodate an additional $SO(3)$ spin group (therefore giving rise
to spin vortices), the corresponding $SO(3)$ symmetry being driven by an
extra scalar sector, see the Ref. \cite{witten,n43}. In view of those
results, Bazeia et al. have recently demonstrated that such an enlarged
Maxwell-Higgs system supports the existence of well-behaved first-order
vortices with internal structures \cite{n44}, whilst arguing that such
solutions may find important applications in the context of metamaterials 
\cite{n45}.

We now go a little bit further into such a subject by studying the
occurrence of first-order rotationally symmetric solitons with internal
structures in a gauged $CP(2)$ model containing an additional scalar field.
We have implemented successfully the Bogomol'nyi prescription by obtaining
an energy lower-bound (Bogomol'nyi limit) and the respective BPS equations
satisfied by the fields saturating that bound. We then solve these equations
by means of a finite-difference scheme, via which we obtain regular vortices
with finite energy. The point to be raised here is that the resulting
configurations differ from their canonical counterparts (obtained in the
absence of the additional field), these differences being understood as
internal structures, as we clarify below.

In order to present our results, this work is organized as follows: in the
Sec. II, we introduce the extended model in which the complex $CP(2)$ field
couples minimally to the Maxwell term and also interacts with an additional
noncharged real scalar field. This additional field also composes the
dielectric function multiplying the Maxwell term and the potential of the
new model. We focus our attention on those time-independent solitons with
rotational symmetry. In this context, we proceed the minimization of the
effective energy, from which we obtain a set of three first-order
differential equations and a well-defined lower bound for the total energy.
In addition, in the Sec. III, we split our investigation into two different
cases based on different choices for the dielectric function. In the sequel,
we solve the corresponding first-order equations numerically, the resulting
profiles dramatically differing from the canonical ones (obtained in the
absence of the additional field). We point out such differences can be
understood as the so-called internal structures. In the Sec. IV, we end our
work by presenting our final comments and perspectives regarding future
investigations.

In this work, we use the natural units system, for the sake of simplicity.


\section{The model \label{2}}

\label{general}

We begin our manuscript by introducing the Lagrange density which defines
the (2+1)-dimensional field model under investigation, i.e.%
\begin{eqnarray}
\mathcal{L} &=&-\frac{1}{4}G(\chi )F_{\mu \nu }F^{\mu \nu }+\left(
P_{ab}D_{\mu }\phi _{b}\right) ^{\ast }P_{ac}D^{\mu }\phi _{c}  \notag \\
&&+\frac{1}{2}\partial _{\mu }\chi \partial ^{\mu }\chi -V(\chi ,\left\vert
\phi _{3}\right\vert )\text{,}  \label{t1}
\end{eqnarray}%
where $F_{\mu \nu }=\partial _{\mu }A_{\nu }-\partial _{\nu }A_{\mu }$ is
the usual electromagnetic field strength tensor and $P_{ab}=\delta
_{ab}-h^{-1}\phi _{a}\phi _{b}^{\ast }$ stands for a projection operator
defined conveniently. Also, the scalar $CP(2)$ field $\phi _{a}\left( x^{\mu
}\right) $\ (with $a=1,2,3$) is constrained to satisfy $\phi _{a}^{\ast
}\phi _{a}=h$. The $CP(2)$ field is minimally coupled to the gauge one via
the covariant derivative defined by%
\begin{equation}
D_{\mu }\phi _{a}=\partial _{\mu }\phi _{a}-igA_{\mu }Q_{ab}\phi _{b}\text{,}
\end{equation}%
where $g$ is the electromagnetic coupling constant and $Q_{ab}$ represents
the charge matrix (real, diagonal and traceless) \cite{loginov}.

Here, it is important to highlight the presence of an additional scalar
field (neutral and real) $\chi \left( x^{\mu }\right) $ in the Lagrange
density (\ref{t1}). This field couples to the gauge sector via an a priori
arbitrary dielectric function $G(\chi )$ multiplying the Maxwell's term. It
is also supposed to interact with the original $CP(2)$ field via the
potential function $V(\chi ,\left\vert \phi _{3}\right\vert )$ which
spontaneously breaks the original $SU(3)$\ symmetry into the $SU(2)$\ one,
as expected (given that topologically nontrivial configurations are known to
be formed as a consequence of such a phase transition).

In fact, as we demonstrate later below, the presence of an additional scalar
field introduces interesting changes in the shape of the final first-order
vortex solutions in comparison to their canonical counterparts already
studied in \cite{casana} (obtained in the absence of such a neutral field).

It is now instructive to write down the Euler-Lagrange equation for the
gauge sector coming from the model (\ref{t1}). It reads%
\begin{equation}
\partial _{\beta }\left( GF^{\lambda \beta }\right) =J^{\lambda }\text{,}
\end{equation}%
where%
\begin{equation}
\frac{J^{\lambda }}{ig}=P_{ac}D^{\lambda }\phi _{c}\left( P_{ab}\right)
^{\ast }Q_{bf}\phi _{f}^{\ast }-\left( P_{ab}D^{\lambda }\phi _{b}\right)
^{\ast }P_{ac}Q_{cb}\phi _{b}
\end{equation}%
stands for the current 4-vector.

In particular, the Gauss law for time-independent fields can be written as
(Latin indices mean summation over spatial coordinates only)%
\begin{equation}
\partial _{i}\left( G\partial ^{i}A^{0}\right) =-J^{0}\text{,}
\end{equation}%
in which%
\begin{equation}
\frac{J^{0}}{ig}=P_{ab}D^{0}\phi _{b}\left( P_{ac}Q_{cd}\phi _{d}\right)
^{\ast }-\left( P_{ab}D^{0}\phi _{b}\right) ^{\ast }P_{ac}Q_{cd}\phi _{d}%
\text{,}
\end{equation}%
with $D^{0}\phi _{b}=-igQ_{bc}\phi _{c}A^{0}$. Here, given that $A^{0}=0$
satisfies the static Gauss law identically, one concludes that the
time-independent solutions the theory (\ref{t1}) supports present no
electric field.

In such a context, we look for vortex configurations via the usual
rotationally symmetric Ansatz:%
\begin{equation}
A_{i}=-A^{i}=-\frac{1}{gr}\epsilon ^{ij}n^{j}A(r)\text{,}  \label{ta1}
\end{equation}%
\begin{equation}
\left( 
\begin{array}{c}
\phi _{1} \\ 
\phi _{2} \\ 
\phi _{3}%
\end{array}%
\right) =h^{\frac{1}{2}}\left( 
\begin{array}{c}
e^{im_{1}\theta }\sin \left( \alpha (r)\right) \cos \left( \beta (r)\right) 
\\ 
e^{im_{2}\theta }\sin \left( \alpha (r)\right) \sin \left( \beta (r)\right) 
\\ 
e^{im_{3}\theta }\cos \left( \alpha (r)\right) 
\end{array}%
\right) \text{,}  \label{ta2}
\end{equation}%
together with $\chi =\chi (r)$. Here, $\epsilon ^{ij}$ (with $\epsilon
^{12}=+1$) and $n^{j}=\left( \cos \theta ,\sin \theta \right) $ are the
bidimensional antisymmetric tensor and the unit vector, respectively. Also, $%
m_{1}$, $m_{2}$ and $m_{3}$ stand for the winding numbers rotulating the
final structures.

The rotationally symmetric magnetic field is given by 
\begin{equation}
B(r) =-\frac{1}{gr}\frac{dA}{dr}.  \label{tmf}
\end{equation}

Now, regarding the combination between the real charge matrix $Q_{ab}$ and
the winding numbers $\left( m_{1},m_{2},m_{3}\right) $, it is already known
that there is only one effective scenario supporting the existence of
well-behaved first-order vortices, see the discussion in \cite{loginov}.
Therefore, in what follows, we choose to work with $m_{1}=-m_{2}=m$ (with $%
m\in \mathbb{Z}$), $m_{3}=0$ and 
\begin{equation}
Q_{ab}=\frac{1}{2}\text{diag}\left( 1,-1,0\right) \text{,}
\end{equation}%
for the sake of simplicity.

In addition, given the choices stated above, one gets two different
solutions for the profile function $\beta (r)$, i.e.%
\begin{equation}
\beta (r)=\beta _{1}=\frac{\pi }{4}+\frac{\pi }{2}k\text{ \ \ or \ \ }\beta
(r)=\beta _{2}=\frac{\pi }{2}k\text{,}  \label{t3}
\end{equation}%
with $k\in \mathbb{Z}$. However, it was also demonstrated in \cite{casana}
that the field equations arising from these two a priori different cases
simply mimic each other, the resulting contexts being then
phenomenologically equivalent (at least regarding topological solitons at
the classical level). In this sense, in the remainder of the present
manuscript, we consider the case $\beta (r)=\beta _{1}$ only.

\subsection{BPS formalism: The case $\displaystyle{\protect\beta (r)=\protect%
\beta _{1}=\frac{\protect\pi }{4}+\frac{\protect\pi }{2}k}$}

As a consequence of the choice $\beta (r)=\beta _{1}$, the profile functions 
$\alpha (r)$ and $A(r)$ are supposed to obey the standard boundary
conditions, i.e.%
\begin{equation}
\alpha \left( r=0\right) =0\text{ \ \ and \ \ }A\left( r=0\right) =0\text{,}
\label{tbc1}
\end{equation}%
\begin{equation}
\alpha \left( r\rightarrow \infty \right) \rightarrow \frac{\pi }{2}\text{ \
\ and \ \ }A\left( r\rightarrow \infty \right) \rightarrow 2m\text{,}
\label{tbc2}
\end{equation}%
which give rise to regular (nonsingular) configurations with finite energy.

It is important to say that all the equations we present from this point on
describe the effective scenario defined by the conventions argued above.

We focus our attention on those time-independent solutions satisfying a
given set of first-order differential equations. Here, we find these
equations by proceeding the implementation of the usual Bogomol'nyi
algorithm, i.e. by minimizing the total energy of the overall system, the
starting-point being the energy-momentum tensor coming from (\ref{t1})%
\begin{eqnarray}
\mathcal{T}_{\lambda \rho } &=&-GF_{\mu \lambda }F^{\mu }{}_{\rho }+\left(
P_{ab}D_{\lambda }\phi _{b}\right) ^{\ast }P_{ac}D_{\rho }\phi _{c}  \notag
\\[0.2cm]
&&+\left( P_{ab}D_{\rho }\phi _{b}\right) ^{\ast }P_{ac}D_{\lambda }\phi
_{c}+\partial _{\lambda }\chi \partial _{\rho }\chi -\eta _{\lambda \rho }%
\mathcal{L}\text{,}\quad
\end{eqnarray}%
where $\eta _{\lambda \rho }=(+--)$ stands for the metric signature of the
flat spacetime.

The energy density $\varepsilon =T_{00}$ can then be written in the form%
\begin{eqnarray}
\varepsilon  &=&-GF_{\mu 0}F^{\mu }{}_{0}+2\left( P_{ab}D_{0}\phi
_{b}\right) ^{\ast }P_{ac}D_{0}\phi _{c}  \notag \\[0.2cm]
&&+\partial _{0}\chi \partial _{0}\chi -\eta _{00}\mathcal{L}\text{,}
\end{eqnarray}%
its rotationally symmetric version reading%
\begin{eqnarray}
\varepsilon  &=&\frac{1}{2}GB^{2}+\frac{1}{2}\left( \frac{d\chi }{dr}\right)
^{2}+V(\chi ,\alpha )  \notag \\
&&+h\left[ \left( \frac{d\alpha }{dr}\right) ^{2}+\left( \frac{A}{2}%
-m\right) ^{2}\frac{\sin ^{2}\alpha }{r^{2}}\right] \text{.}  \label{te1}
\end{eqnarray}

After some algebraic manipulations, the energy density (\ref{te1}) can be
rewritten as%
\begin{eqnarray}
\varepsilon  &=&\frac{1}{2}G\left( B\mp \sqrt{\frac{2U}{G}}\right) +h\left[ 
\frac{d\alpha }{dr}\mp \left( \frac{A}{2}-m\right) \frac{\sin \alpha }{r}%
\right] ^{2}  \notag \\[0.2cm]
&&+\frac{1}{2}\left( \frac{d\chi }{dr}\mp \frac{1}{r}\frac{d\mathcal{W}}{%
d\chi }\right) ^{2}+V-U-\frac{1}{2r^{2}}\left( \frac{d\mathcal{W}}{d\chi }%
\right) ^{2}  \notag \\[0.2cm]
&&\pm B\sqrt{2GU}\pm h\left( A-2m\right) \frac{\sin \alpha }{r}\frac{d\alpha 
}{dr}\pm \frac{1}{r}\frac{d\mathcal{W}}{dr}\text{,}  \label{te1ab}
\end{eqnarray}%
where we have introduced both the nonnegative function $U\equiv U(\chi
,\alpha )$ and $W\equiv W(\chi )$.

Now, given the expression (\ref{tmf}) for the magnetic field, it is possible
to transform the first and second terms of the third row in (\ref{te1ab}) in
a total derivative by imposing the constraint%
\begin{equation}
\frac{d}{dr}\left( \sqrt{2GU}\right) =gh\frac{d}{dr}\left( \cos \alpha
\right) \text{,}  \label{tkc}
\end{equation}%
which leads to a relation between $U\left( \chi ,\alpha \right) $ and the
dielectric function $G\left( \chi \right) $, i.e.%
\begin{equation}
U\left( \chi ,\alpha \right) =\frac{g^{2}h^{2}}{2G(\chi )}\cos ^{2}\alpha 
\text{,}  \label{tv1}
\end{equation}%
where we have chosen a null integration constant.

In this sense, by putting all the previous results in (\ref{te1ab}), we get%
\begin{eqnarray}
\varepsilon  &=&\frac{1}{2}G\left( B\mp \frac{gh}{G}\cos \alpha \right) +h%
\left[ \frac{d\alpha }{dr}\mp \left( \frac{A}{2}-m\right) \frac{\sin \alpha 
}{r}\right] ^{2}  \notag \\[0.2cm]
&&\hspace{-0.25cm}+\frac{1}{2}\left( \frac{d\chi }{dr}\mp \frac{1}{r}\frac{d%
\mathcal{W}}{d\chi }\right) ^{2}\mp \frac{1}{r}\frac{d}{dr}\left[ \frac{{}}{%
{}}\!h\left( A-2m\right) \cos \alpha -\mathcal{W}\right]   \notag \\[0.2cm]
&&\hspace{-0.25cm}+V-\frac{g^{2}h^{2}}{2G(\chi )}\cos ^{2}\alpha -\frac{1}{%
2r^{2}}\left( \frac{d\mathcal{W}}{d\chi }\right) ^{2}\text{,}  \label{te1a}
\end{eqnarray}%
from which we immediately choose the potential to be%
\begin{equation}
V\left( \chi ,\alpha \right) =\frac{g^{2}h^{2}}{2G(\chi )}\cos ^{2}\alpha +%
\frac{1}{2r^{2}}\left( \frac{d\mathcal{W}}{d\chi }\right) ^{2}\text{,}
\label{tp}
\end{equation}%
or%
\begin{equation}
V\left( \chi ,\left\vert \phi _{3}\right\vert \right) =\frac{g^{2}h}{2G(\chi
)}\left\vert \phi _{3}\right\vert ^{2}+\frac{1}{2r^{2}}\left( \frac{d%
\mathcal{W}}{d\chi }\right) ^{2}\text{,}
\end{equation}%
when written in terms of $\left\vert \phi _{3}\right\vert $.

We have therefore obtained that the rotationally symmetric expression for
the energy distribution can be written according the Bogomol'nyi idea, from
which the Eq. (\ref{te1a}) becomes%
\begin{eqnarray}
\varepsilon  &=&\varepsilon _{bps}+h\left[ \frac{d\alpha }{dr}\mp \frac{1}{r}%
\left( \frac{A}{2}-m\right) \sin \alpha \right] ^{2}  \notag \\[0.2cm]
&&+\frac{1}{2}G\left( B\mp \frac{gh}{G}\cos \alpha \right) ^{2}+\frac{1}{2}%
\left( \frac{d\chi }{dr}\mp \frac{1}{r}\mathcal{W}_{\chi }\right) ^{2}\text{,%
}\hspace{0.5cm}  \label{ted}
\end{eqnarray}%
where 
\begin{equation}
\varepsilon _{bps}=\mp \frac{1}{r}\frac{d}{dr}\left[ h\left( A-2m\right)
\cos \alpha -\mathcal{W}\right] \text{.}
\end{equation}

It is important to comment about the explicit dependence of the potential $%
V\left( \chi , \phi \right) $ on the radial coordinate $r$ (the last term in
(\ref{tp})). Indeed, such a term should not represent a dramatic novelty: it
was already considered in \cite{20} in order to successfully circumvent the
Derrick-Hobart theorem \cite{2122}; it was also used in \cite{n44} itself in
order to study those first-order solitons with internal structures arising
from the simplest Maxwell-Higgs scenario.

We now return to the expression for the energy density in (\ref{ted}), via
which we write the corresponding total energy as%
\begin{equation}
E=2\pi \int_{0}^{\infty }r\varepsilon (r)dr=E_{bps}+{E_{1}}\geq E_{bps}\text{%
,}  \label{tte}
\end{equation}%
where $E_{bps}$ defining the energy lower-bound (i.e. the Bogomol'nyi one)
is given by%
\begin{eqnarray}
E_{bps} &=&2\pi \int_{0}^{\infty }\varepsilon _{bps}rdr  \notag \\[0.2cm]
&=&\mp 2\pi \left[ 2hm+\mathcal{W}(0)-\mathcal{W}(\infty )\right] \text{,}
\label{te}
\end{eqnarray}%
which is always positive whether we consider the lower (upper) sign for $%
m>0\;(m<0)$ and $W(0)>W(\infty )\;(W(0)<W(\infty ))$. Also, the term $E_{1}$
is given by%
\begin{eqnarray}
E_{1} &=&2\pi \int_{0}^{\infty }\left\{ h\left[ \frac{d\alpha }{dr}\mp \frac{%
1}{r}\left( \frac{A}{2}-m\right) \sin \alpha \right] ^{2}\right.   \notag \\
&&\hspace{1.5cm}+\frac{1}{2}G\left( B\mp \frac{gh}{G}\cos \alpha \right) ^{2}
\notag \\
&&\hspace{1.5cm}\left. +\frac{1}{2}\left( \frac{d\chi }{dr}\mp \frac{1}{r}%
\mathcal{W}_{\chi }\right) ^{2}\right\} rdr\text{.}
\end{eqnarray}

Now, from the total energy as it appears in (\ref{tte}), one concludes
that,\ whether $E_{1}=0$, i.e. if the fields satisfy the first-order
equations (the Bogomol'nyi ones)%
\begin{equation}
\frac{d\alpha }{dr}=\pm \frac{1}{r}\left( \frac{A}{2}-m\right) \sin \alpha 
\text{,}  \label{tbps2}
\end{equation}%
\begin{equation}
-\frac{1}{gr}\frac{dA}{dr}=\pm \frac{gh}{G}\cos \alpha \text{,}
\label{tbps1}
\end{equation}%
\begin{equation}
\frac{d\chi }{dr}=\pm \frac{1}{r}\mathcal{W}_{\chi }\text{,}  \label{tnf}
\end{equation}%
the resulting rotationally symmetric configurations saturate the
well-defined lower bound (\ref{te}) for the total energy (in equation (\ref%
{tbps1}), we have used (\ref{tmf}) to represent the magnetic field). 

We summarize the resulting scenario as follows: given a potential (\ref{tp})
constructed by choosing adequately the dielectric function $G(\chi )$ and
the superpotential $W(\chi )$, the rotationally symmetric fields $\alpha (r)$%
, $A(r)$ and $\chi (r)$ satisfy the Bogomol'nyi equations (\ref{tbps2}), (%
\ref{tbps1}) and (\ref{tnf}), therefore giving rise to time-independent
nonsingular configurations with total energy equal to (\ref{te}).

In addition, it is important to observe that the solution of the equation (%
\ref{tnf}) depends on the form of the superpotential $W(\chi )$. Despite of
the fact that the Eq. (\ref{tnf}) seems uncoupled of the other two BPS
equations, its solution $\chi (r)$ affects the ones for both $A(r)$ and $%
\alpha (r)$ via the dielectric function $G(\chi )$. In fact, as we
demonstrate in the next Section, such influence introduces significant
changes in the internal structure of the first-order vortices that the model
(\ref{t1}) supports.


\section{First-order solutions with internal structures}

It is interesting to highlight that in the absence of the additional field $%
\chi (r) $ and for $G(\chi)=1$, the model (\ref{t1}) reduces to the gauged $%
CP(2)$ one whose first-order solitons were studied in \cite{casana}. On the
other hand, when both the gauge and scalar $CP(2)$ sectors vanish, the
theory (\ref{t1}) leads us back to the scalar self-dual scenario considered
in \cite{20}.

We now return to the complete model\ (\ref{t1}), for which we proceed the
explicit construction of those BPS vortices presenting internal structures
by using the first-order equations introduced in the previous Section.

We begin to solve the BPS scenario through the first-order Eq. (\ref{tnf})
for the real field $\chi $. In this sense, we choose the superpotential $%
W(\chi )$ as%
\begin{equation}
\mathcal{W}(\chi )=\chi -\frac{1}{3}\chi ^{3}\text{,}  \label{tw1}
\end{equation}%
which was previously used in \cite{2324} in order to study bidimensional
skyrmion-like solitons, and also in \cite{25} as an attempt to understand
the behavior of massless Dirac fermions in a planar skyrmion-like background.

Then, given the expression in (\ref{tw1}), one gets that the first-order
equation (\ref{tnf}) reduces to%
\begin{equation}
\frac{d\chi }{dr}=\pm \frac{1}{r}\left( 1-\chi ^{2}\right) \text{,}
\end{equation}%
whose exact solution is given by%
\begin{equation}
\chi (r) =\pm \frac{r^{2}-r_{0}^{2}}{r^{2}+r_{0}^{2}}\text{,}  \label{tx}
\end{equation}%
where $r_{0}$ stands for an arbitrary positive constant such that $\chi
\left( r=r_{0}\right) =0$. Here, it is important to point out that solution (%
\ref{tx}) satisfy $\chi \left( r=0\right) =\mp 1$ and $\chi \left(
r\rightarrow \infty \right) \rightarrow \pm 1$.

With this information in mind, we calculate the BPS energy (\ref{te}), i.e.%
\begin{equation}
E=E_{bps}=4\pi h\left\vert m\right\vert +\frac{8\pi }{3}\text{,}  \label{teb}
\end{equation}%
which stands for the total energy of the first-order configurations we
introduce below (i.e. the effective Bogolmol'nyi bound).

In the sequel, we split our investigation into two different cases based on
the mathematical forms we choose for the dielectric function $G(\chi )$.

\subsection{The first case}

We firstly select the dielectric function $G\left( \chi \right) $ as%
\begin{equation}
G\left( \chi \right) =\frac{1}{1-\chi ^{2}}\text{,}  \label{tg1}
\end{equation}%
its relevance lying on the fact that, for $r=r_{0}$, the additional field $%
\chi (r) $ vanishes and the dielectric function $G$ equals the unity: in
this case, as we have argued in the beginning of the present Section, the
resulting solutions mimic those ones previously studied in the context of a
purely $CP(2)$ model in the presence of the Maxwell term \cite{casana}. On
the other hand, for $r\neq r_{0}$, the presence of a nontrivial solution for 
$\chi (r) $ given by (\ref{tx})\ is\ expected to change the shape of the
corresponding first-order\ vortices in a new way.

\begin{figure}[tbp]
\centering\includegraphics[width=8.5cm]{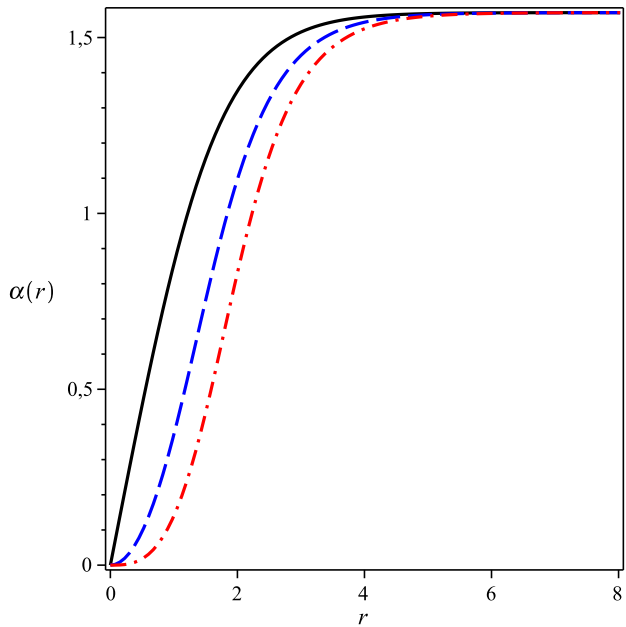}
\par
\vspace{-0.3cm}
\caption{Numerical solutions to $\protect\alpha (r) $ arising from (\protect
\ref{tbps1a}) and (\protect\ref{tbps1b}) via (\protect\ref{tbc1}) and (%
\protect\ref{tbc2}) for $m=1$ (solid black line), $m=2$ (dashed blue line)
and $m=3$ (dash-dotted red line). Here, we have chosen $h=1$, $g=2 $ and $%
r_{0}=5$.}
\end{figure}

Moreover, for $r=0$ and $r\rightarrow \infty $, the term $\chi ^{2}$ equals
the unity (see the discussion after the Eq. (\ref{tx})) and therefore
dielectric function (\ref{tg1}) diverges. However, such a divergence is
compensated by a convenient behavior of the magnetic field $B(r)$, which
avoids the first term in the right-hand side of the expression (\ref{te1})
for the energy density to be singular. In this case, the overall energy
results in the finite value already calculated in (\ref{teb}).

Now, in view of the exact solution (\ref{tx}), the dielectric function (\ref%
{tg1}) can be written in the form%
\begin{equation}
G(r)=\frac{\left( r^{2}+r_{0}^{2}\right) ^{2}}{4r^{2}r_{0}^{2}}\text{,}
\end{equation}%
the first-order BPS equations\ (\ref{tbps2}) and (\ref{tbps1}) for the $CP(2)
$ sector standing for%
\begin{equation}
\frac{d\alpha }{dr}=\pm \frac{1}{r}\left( \frac{A}{2}-m\right) \sin \alpha 
\text{,}  \label{tbps1b}
\end{equation}%
\begin{equation}
\frac{1}{r}\frac{dA}{dr}=\mp \frac{4r_{0}^{2}g^{2}h}{\left(
r^{2}+r_{0}^{2}\right) ^{2}}r^{2}\cos \alpha \text{.}  \label{tbps1a}
\end{equation}%
These Bogomol'nyi equations must be solved numerically by means of a
finite-difference scheme according the boundary conditions (\ref{tbc1}) and (%
\ref{tbc2}).

\begin{figure}[tbp]
\centering\includegraphics[width=8.5cm]{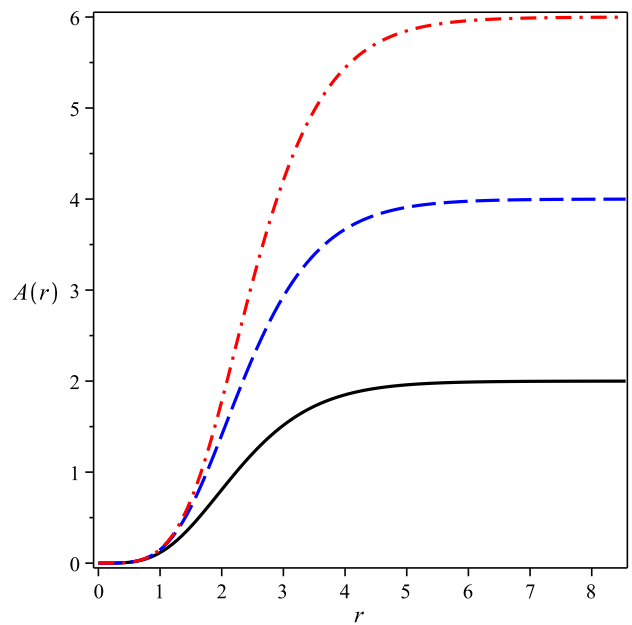}\vspace{-0.3cm}
\caption{Numerical solutions to $A(r) $. Conventions as in the Fig. 1. The
solution approaches the boundary values in a monotonic way, as expected.}
\end{figure}

We now verify the way the profiles $\alpha (r)$ and $A(r)$ approximate the
boundary values. For such an analysis, we consider $m>0$ (i.e. the lower
signs in the first-order equations), for the sake of simplicity. In this
case, near the origin, we represent the profile fields by%
\begin{equation}
\alpha (r)\approx \delta \alpha (r)\text{ \ \ and \ \ }A(r)\approx \delta
A(r)\text{,}  \label{bhvzero}
\end{equation}%
where $\delta \alpha (r)$ and $\delta A(r)$ are small fluctuations around
the boundary values. Now, substituting these representations into the
first-order equations (\ref{tbps1b}) and (\ref{tbps1a}), and taking into
account only the relevant contributions, one gets%
\begin{equation}
\frac{d}{dr}\delta \alpha =m\frac{\delta \alpha }{r}\text{,}  \label{tbpsy}
\end{equation}%
\begin{equation}
\frac{d}{dr}\delta A=\frac{4g^{2}h}{r_{0}^{2}}r^{3}\text{,}  \label{tbpsx}
\end{equation}%
whose solutions provide the behavior of $\alpha (r)$ and $A(r)$, i.e.%
\begin{equation}
\alpha (r)\approx C_{0}r^{m}\text{,}  \label{tas1}
\end{equation}%
\begin{equation}
A(r)\approx \frac{g^{2}h}{r_{0}^{2}}r^{4}\text{,}  \label{tas2}
\end{equation}%
where $C_{0}$ stands for a positive real constant to be determined
numerically.

\begin{figure}[tbp]
\centering\includegraphics[width=8.5cm]{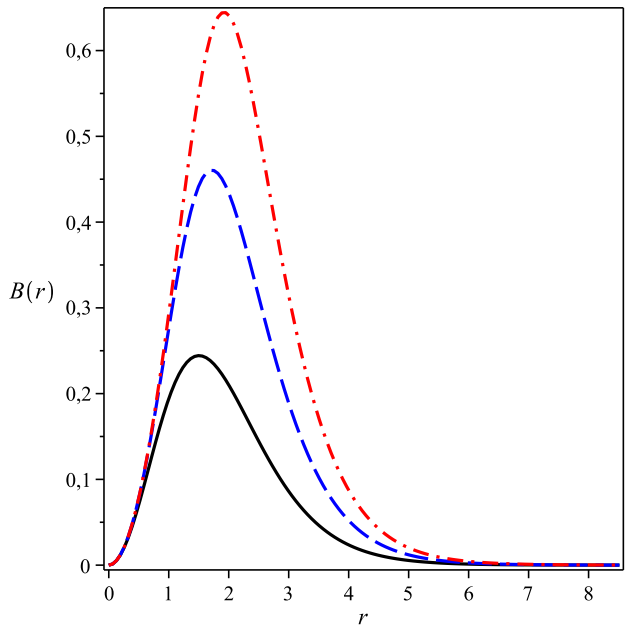}\vspace{-0.3cm}
\caption{Numerical solutions to the modulus of the magnetic field $B(r)$.
Conventions as in the Fig. 1. Here, even in the presence of topological
conditions, the magnetic field mimics the nontopological behavior due to the
nontrivial profile to the extra field $\protect\chi (r)$.}
\end{figure}

A similar procedure can be implemented in order to study the behavior of the
profile fields in the asymptotic limit $r\rightarrow \infty $. In this
sense, we now represent these fields by%
\begin{equation}
\alpha (r)\approx \frac{\pi }{2}-\delta \alpha (r)\text{ \ \ and \ \ }%
A(r)\approx 2m-\delta A(r)\text{,}  \label{bhvinfty}
\end{equation}%
from which, again taking into account only the relevant contributions, one
gets%
\begin{equation}
\frac{d}{dr}\delta \alpha =-\frac{\delta A}{2r}\text{,}
\end{equation}%
\begin{equation}
\frac{d}{dr}\delta A=-4r_{0}^{2}g^{2}h\frac{\delta \alpha }{r}\text{,}
\end{equation}%
whose solutions allow us to conclude that the profile functions themselves
behave as%
\begin{equation}
\alpha (r)\approx \frac{\pi }{2}-C_{\infty }r^{-\sqrt{2h}r_{0}g}\text{,}
\label{tas3}
\end{equation}%
\begin{equation}
A(r)\approx 2m-2\sqrt{2h}r_{0}gC_{\infty }r^{-\sqrt{2h}r_{0}g}\text{.}
\label{tas4}
\end{equation}%
where $C_{\infty }$ represents an integration constant to be also determined
numerically.

We now plot the numerical solutions we have found via the first-order
equations (\ref{tbps1b}) and (\ref{tbps1a}) using the boundary conditions (%
\ref{tbc1}) and (\ref{tbc2}). Here, we have fixed $h=1$, $g=2$ and $r_{0}=5$%
, from which we have solved the equations for $m=1$ (solid black line), $m=2$
(dashed blue line) and $m=3$ (dash-dotted red line).

The figures 1 and 2 show, respectively, the numerical profiles to the $CP(2)$
scalar and gauge functions, i.e. $\alpha (r) $ and $A\left( r\right) $, from
which we see that both solutions exhibit a well-defined monotonic behavior,
approaching the boundary values according the approximate analytical
solutions written in (\ref{tas1}), (\ref{tas2}), (\ref{tas3}) and (\ref{tas4}%
).

\begin{figure}[tbp]
\centering\includegraphics[width=8.5cm]{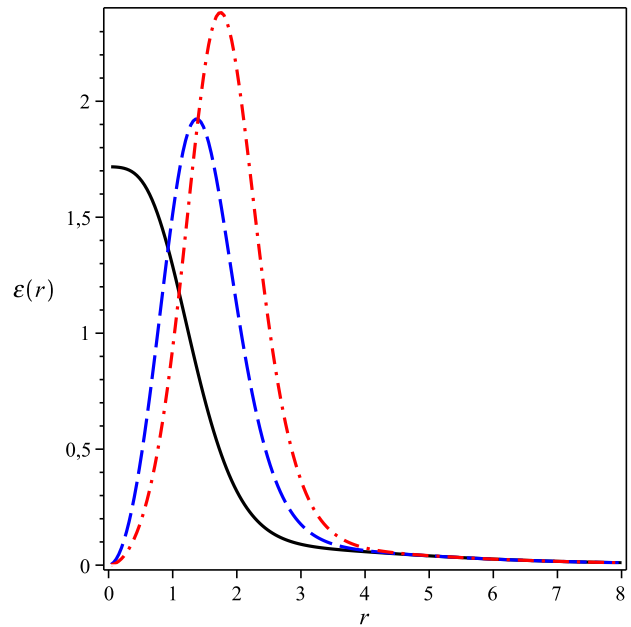}
\par
\vspace{-0.3cm}
\caption{Numerical solutions to the energy density $\protect\varepsilon (r)$%
. Conventions as in the Fig. 1. In this case, for $m=1$, one gets $\protect%
\varepsilon _{bps}\left( r=0\right) =2h\left( C_{0}\right) ^{2}$.}
\end{figure}

In the Figure 3, we present the numerical solutions to the modulus of the
magnetic field $B(r)$. In this case, it is interesting to note how a
nontrivial profile to the function $\chi (r) $ changes the shape of the
corresponding magnetic sector: even in the presence of the standard
topological conditions (\ref{tbc1}) and (\ref{tbc2}), the resulting magnetic
field stands for a ring centered at the origin (in a dramatic contrast with
its usual counterpart, which represents a lump centered at $r=0$; see the
Fig. 2 of the Ref. \cite{casana}), therefore mimicking the typical
nontopological behavior (see the Fig. 3 of the Ref. \cite{lima}, for
instance).

Moreover, given the approximate solutions (\ref{tas2}) and (\ref{tas4}) to
the profile function $A(r) $, one gets that, near the origin, the magnetic
field (\ref{tmf}) behaves as%
\begin{equation}
B(r) \approx -\frac{4g^{2}h}{r_{0}^{2}}r^{2}\text{,}
\end{equation}%
its asymptotic solution reading%
\begin{equation}
B(r) \approx -4ghr_{0}^{2}C_{\infty }r^{-\left( \sqrt{2h}r_{0}g+2\right) }%
\text{,}
\end{equation}%
which confirm that both $B\left( r=0\right) $ and $B\left( r\rightarrow
\infty \right) $ vanish, in agreement to the Figure 3.

\begin{figure}[tbp]
\centering\includegraphics[width=8.5cm]{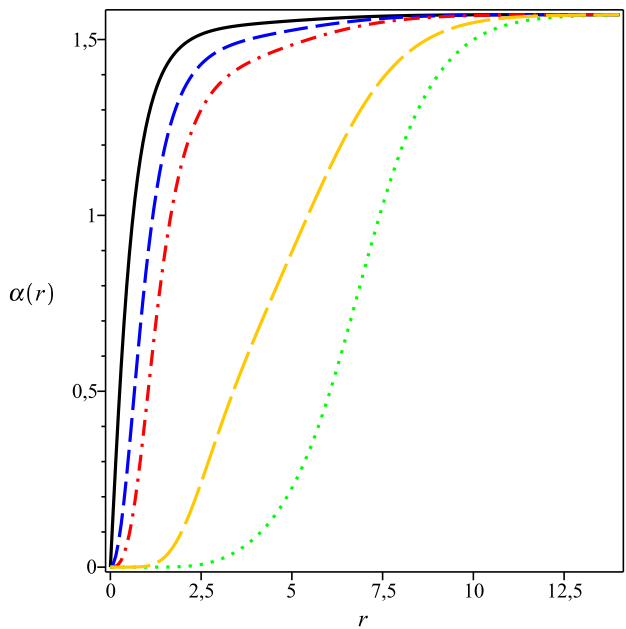}
\par
\vspace{-0.3cm}
\caption{Numerical solutions to $\protect\alpha (r) $ arising from (\protect
\ref{tbps2a}) and (\protect\ref{tbps2b}) via (\protect\ref{tbc1}) and (%
\protect\ref{tbc2}). Conventions as in the Fig. 1. In addition, we also
depict the results for $m=7$ (long-dashed orange line) and $m=10$ (dotted
green line).}
\end{figure}

The Figure 4 presents the numerical profiles to the energy density $%
\varepsilon _{bps}(r) $. Here, it is worthwhile to point out that, for $m=1$%
, the energy distribution stands for a lump centered at the origin
(therefore mimicking the canonical result, see the Fig. 4 of the Ref. \cite%
{casana}). In particular, in view of the approximate expressions (\ref{tas1}%
) and (\ref{tas2}), one gets that, near $r=0$, the energy density obeys%
\begin{equation}
\varepsilon _{bps}(r) \approx 2h\left( C_{0}\right) ^{2}+\frac{4}{r_{0}^{2}}%
\left( \frac{4}{r_{0}^{2}}+g^{4}h^{2}-\frac{h\left( C_{0}\right) ^{4}}{6}%
r_{0}^{2}\right) r^{2}\text{,}
\end{equation}%
from which one additionally concludes that $\varepsilon _{bps}\left(
r=0\right) =2h\left( C_{0}\right) ^{2}$, for $m=1$.

On the other hand, for $m\neq 1$, the final configuration corresponds to a
ring which, near the origin, behaves as%
\begin{equation}
\varepsilon _{bps}(r) \approx \frac{4}{r_{0}^{2}}\left( \frac{4}{r_{0}^{2}}%
+g^{4}h^{2}\right) r^{2}\text{,}
\end{equation}%
therefore vanishing at the origin, its amplitude increasing as the vorticity 
$m$ itself increases. In this case, keeping the standard profile in mind,
one also finds an interesting difference: in the absence of the additional
terms, the energy density's ring does not vanish at $r=0$, its amplitude
decreasing as $m$ increases, see the Fig. 3 of the Ref. \cite{casana}.

\begin{figure}[tbp]
\centering\includegraphics[width=8.5cm]{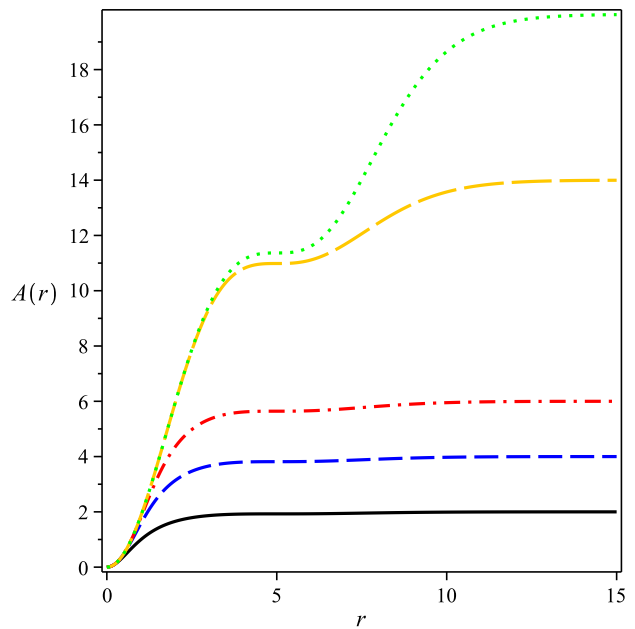} \vspace{-0.3cm}
\caption{Numerical solutions to $A(r) $. Conventions as in the Fig. 5. In
this case, as $m$ increases, $A(r) $ engenders the formation of a plateau
around $r=r_{0}$.}
\end{figure}

We conclude this part by arguing that, given the solutions (\ref{tas3}) and (%
\ref{tas4}), the asymptotic one for the energy density can be verified to
read%
\begin{equation}
\varepsilon _{bps}(r) \approx \frac{16r_{0}^{4}}{r^{6}}\text{,}  \label{taed}
\end{equation}%
which saturates $\varepsilon _{bps}\left( r\rightarrow \infty \right)
\rightarrow 0$ (i.e. the finite energy requirement), therefore confirming
the convenience of the finite energy boundary conditions (\ref{tbc2}).

\subsection{The second case}

We now choose the dielectric function as%
\begin{equation}
G\left( \chi \right) =\frac{1}{\chi ^{2}}\text{.}  \label{tg2}
\end{equation}%
In this case, given that $\chi ^{2}$ calculated from (\ref{tx}) equals the
unity at $r=0$ and $r\rightarrow \infty $, it follows from (\ref{tg2}) that $%
G\left( r=0\right) =1$ and $G\left( r\rightarrow \infty \right) \rightarrow
1 $. As a consequence, at the origin and asymptotically, the corresponding
rotationally symmetric solitons behave as the ones arising from a simplest
gauged $CP(2)$ scenario. Furthermore, at the point $r=r_{0}$, the field $%
\chi $ in (\ref{tx}) vanishes and the function $G$ in (\ref{tg2}) diverges:
in this context, such a divergence is counterbalanced by a null magnetic
field at $r=r_{0}$ (therefore introducing an internal structure to the
resulting vortices), from which one gets that the first term in (\ref{te1})
results nonsingular, the total energy then converging to the value in (\ref%
{teb}).

\begin{figure}[tbp]
\centering\includegraphics[width=8.5cm]{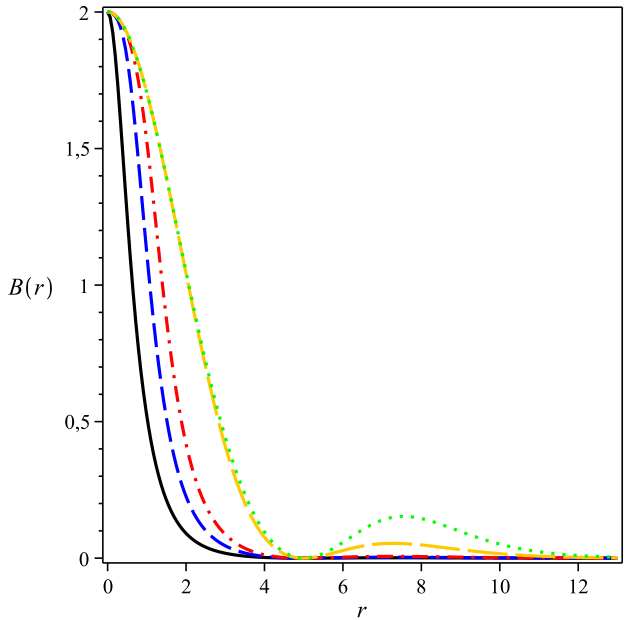} \vspace{-0.3cm}
\caption{Numerical solutions to the modulus of the magnetic field $B(r)$.
Conventions as in the Fig. 5. The magnetic sector vanishes at $r=r_{0}$,
which prevents the corresponding energy distribution to be divergent.}
\end{figure}

We now use (\ref{tx}) in order to write (\ref{tg2}) as%
\begin{equation}
G(r)=\frac{\left( r^{2}+r_{0}^{2}\right) ^{2}}{\left( r^{2}-r_{0}^{2}\right)
^{2}}\text{,}
\end{equation}%
the equations (\ref{tbps2}) and (\ref{tbps1}) therefore reducing to%
\begin{equation}
\frac{d\alpha }{dr}=\pm \frac{1}{r}\left( \frac{A}{2}-m\right) \sin \alpha 
\text{,}  \label{tbps2b}
\end{equation}%
\begin{equation}
\frac{1}{r}\frac{dA}{dr}=\mp \frac{\left( r^{2}-r_{0}^{2}\right) ^{2}}{%
\left( r^{2}+r_{0}^{2}\right) ^{2}}g^{2}h\cos \alpha \text{,}  \label{tbps2a}
\end{equation}%
These equations must be also solved numerically via the conditions (\ref%
{tbc1}) and (\ref{tbc2}).

We proceed to the study of the differential equations (\ref{tbps2b}) and (%
\ref{tbps2a}) near the boundaries, again for $m>0$. We begin by considering,
near the origin, the profiles as represented in (\ref{bhvzero}), the
first-order equations satisfied by the small functions $\delta \alpha $ and $%
\delta A$ reading%
\begin{equation}
\frac{d}{dr}\delta \alpha =m\frac{\delta \alpha }{r}\text{,}
\end{equation}%
\begin{equation}
\frac{d}{dr}\delta A=g^{2}hr-\frac{4g^{2}h}{r_{0}^{2}}r^{3}\text{,}
\end{equation}%
from which one gets the solutions%
\begin{equation}
\alpha (r)\approx \mathcal{C}_{0}r^{m}\text{,}
\end{equation}%
\begin{equation}
A(r)\approx \frac{g^{2}h}{2}r^{2}-\frac{g^{2}h}{r_{0}^{2}}r^{4}\text{,}
\label{ta0}
\end{equation}%
in which the real positive constant $C_{0}$ must be determined numerically. 

Likewise, in the limit $r\rightarrow \infty $, we proceed the implementation
of the profile representation (\ref{bhvinfty}) into the BPS equations (\ref%
{tbps2b}) and (\ref{tbps2a}), from which we obtain the following linearized
equations for the small functions $\delta \alpha $ and $\delta A$,%
\begin{equation}
\frac{d}{dr}\delta \alpha =-\frac{\delta A}{2r}\text{,}
\end{equation}%
\begin{equation}
\frac{1}{r}\frac{d}{dr}\delta A=-g^{2}h\delta \alpha \text{,}
\end{equation}%
whose solutions allow us to write the behavior of the functions $\alpha (r)$
and $A(r)$ as%
\begin{equation}
\alpha (r)\approx \frac{\pi }{2}-\mathcal{C}_{\infty }e^{-Mr}\text{,}
\label{tc2}
\end{equation}%
\begin{equation}
A(r)\approx 2m-g\sqrt{2h}\mathcal{C}_{\infty }re^{-Mr}\text{,}  \label{yc3}
\end{equation}%
where $C_{\infty }$ is an integration constant and $M$ is real parameter
given by%
\begin{equation}
M=g\sqrt{\frac{h}{2}}\text{.}  \label{tc1}
\end{equation}%
In this case, the exponential behavior of the asymptotic solutions (\ref{tc2}%
) and (\ref{yc3}) reveals that the real constant $M$ defined by (\ref{tc1})
in terms of $g$ and $h$ stands for the masses of the $CP(2)$ scalar and
gauge bosons inherent to the original model.

In the sequel, we depict the numerical profiles we have obtained from the
first-order equations (\ref{tbps2b}) and (\ref{tbps2a}) via the boundary
conditions (\ref{tbc1}) and (\ref{tbc2}). Here, we have chosen the same
values as before for the parameters appearing in the first-order equations,
i.e. $h=1$, $g=2$ and $r_{0}=5$, via which we have studied the equations for 
$m=1$ (solid black line), $m=2$ (dashed blue line), $m=3$ (dash-dotted red
line), $m=7$ (long-dashed orange line) and $m=10$ (dotted green line).

\begin{figure}[tbp]
\centering\includegraphics[width=8.5cm]{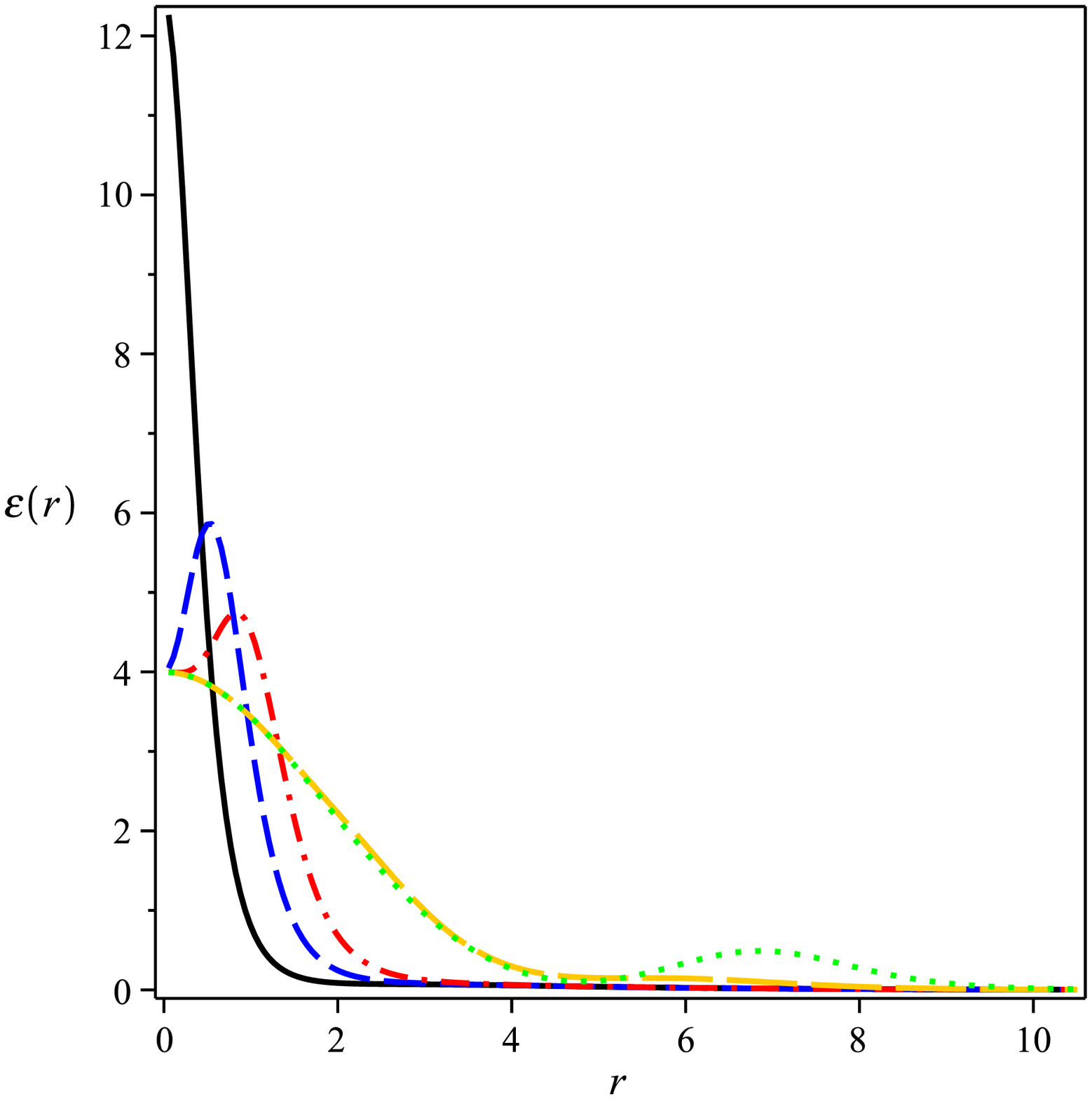} \vspace{-0.3cm}
\caption{Numerical solutions to the energy density $\protect\varepsilon (r)$%
. Conventions as in the Fig. 5. In this case, one gets $\protect\varepsilon %
_{bps}\left( r=0\right) =g^{2}h^{2}+2h\left( \mathcal{C}_{0}\right) ^{2}$,
for $m=1$. }
\end{figure}

In the figures 5 and 6, we present the solutions to the profile functions $%
\alpha (r) $ and $A(r) $, respectively. Here, it is interesting to note how
the shape of the solutions depend on the values of the vorticity $m$. In
particular, as $m$ increases, the field $A\left( r\right) $ engenders the
formation of a plateau around $r=r_{0}$ (it is instructive to compare such
an effect with the one identified in the Fig. 5 of the Ref. \cite{n44}),
such an internal structure modifying the resulting magnetic field's profile.

The Figure 7 shows the solutions to the modulus of the magnetic field $B(r) $%
, the corresponding configuration standing for\ a double-lump centered at
the origin. Here, as we have argued previously, the magnetic sector vanishes
at $r=r_{0}$ (such an effect being a consequence of the plateau appearing in
the solution for $A(r) $), which prevents the first term in the energy
distribution (\ref{te1}) to be divergent, the total energy therefore
converging to the well-defined value in (\ref{teb}) (we highlight that a
similar effect is shown in the Fig. 5 of the Ref. \cite{n44}).

In addition, in view of the approximate solutions (\ref{ta0}) and (\ref{yc3}%
), it can be verified that, near $r=0$, the magnetic sector is given by%
\begin{equation}
B(r) \approx -gh+\frac{4gh}{r_{0}^{2}}r^{2}\text{,}
\end{equation}%
its asymptotic counterpart standing for%
\begin{equation}
B(r) \approx - gh\mathcal{C}_{\infty } e^{-Mr}\text{,}
\end{equation}%
from which one gets that $B\left( r=0\right) =-gh$ and $B\left( r\rightarrow
\infty \right) \rightarrow 0$, in agreement to the Fig. 7.

We end this Section by discussing the numerical profiles to the energy
density $\varepsilon _{bps}(r) $ appearing in the Figure 8. In this case,
for $m=1$, the resulting configuration is a regular lump whose approximate
solution, near the origin, reads%
\begin{equation}
\varepsilon _{bps}(r) \approx g^{2}h^{2}+2h\left( \mathcal{C}_{0}\right)
^{2}+\mathcal{C}_{2}r^{2}\text{,}
\end{equation}%
the real parameter $\mathcal{C}_{2}$ being given by%
\begin{equation}
\mathcal{C}_{2}=2\left( \frac{8}{r_{0}^{4}}-\frac{h}{3}\left( \mathcal{C}%
_{0}\right) ^{4}-g^{2}h^{2}\left( \frac{2}{r_{0}^{2}}+\left( \mathcal{C}%
_{0}\right) ^{2}\right) \right) \text{,}
\end{equation}%
via which we conclude that $\varepsilon _{bps}\left( r=0\right)
=g^{2}h^{2}+2h\left( \mathcal{C}_{0}\right) ^{2}$, for $m=1$.

Moreover, for intermediate values of the vorticity, the solutions stand for
well-behaved rings centered at $r=0$. However, as $m$ increases, these
intermediate rings transmute to nonsingular double-lumps whose intersection
occurs around $r=r_{0}$ (at this point, the energy distribution vanishes).
Here, it is important to highlight that such a transmutation seems to be a
new phenomenon inherent to the gauged $CP(2)$ scenario presently considered.

Finally, the reader can verify that, for $m\neq 1$, the energy distribution,
near the origin, behaves as%
\begin{equation}
\varepsilon _{bps}(r) \approx g^{2}h^{2}+\frac{4}{r_{0}^{2}}\left( \frac{4}{%
r_{0}^{2}}-g^{2}h^{2}\right) r^{2}\text{,}
\end{equation}%
which gives $\varepsilon _{bps}\left( r=0\right) =g^{2}h^{2}$. In addition,
for any value of the vorticity $m$, the asymptotic solution for the energy
density can be verified to be the very same one appearing in (\ref{taed}),
therefore satisfying the finite energy requirement and the convenience of
the conditions in (\ref{tbc2}).


\section{Final comments}

We have considered a gauged $CP(2)$ model containing an additional real
scalar field which couples to the electromagnetic one via a dielectric
function multiplying the usual Maxwell's term. The stationary Gauss law
tells us that the time-independent configurations have no electric charge.
We have particularized our investigation by choosing well-established
conventions supporting the existence of rotationally symmetric solitons with
finite energy.

We then have developed the corresponding first-order framework by means of
the Bogomol'nyi prescription (i.e. by minimizing the total energy of the
effective model). Consequently, we have obtained the corresponding set of
first-order differential equations (the Bogomol'nyi ones) engendering
genuine field solutions saturating a lower-bound for the resulting energy
(the Bogomol'nyi bound). It is interesting to point out that the first-order
equations depend on two a priori arbitrary functions, i.e. the
superpotential $\mathcal{W}(\chi)$ for the additional field $\chi $ and the
dielectric function $G(\chi)$. In view of such a dependence, we have
splitted our investigation into two different cases based on different
choices for the dielectric function $G (\chi ) $.

We have then explored such a freedom in order to construct regular
first-order vortices whose shapes dramatically differ from their canonical
counterparts (obtained in the absence of the additional field), the new
details being understood as internal structures, as argued in the previous
work \cite{n44}.

It is important to say that the results we have presented in this manuscript
hold a priori only for those rotationally symmetric configurations defined
by the ansatz (\ref{ta1}) and (\ref{ta2}), being therefore not possible to
ensure that the original theory supports such a first-order framework
outside the rotationally symmetric scenario, such question lying beyond the
scope of this work.

Interesting ideas regarding future investigations include the search for
electrically charged first-order vortices with internal structures arising
from both the Chern-Simons and the Maxwell-Chern-Simons versions of the
original model (\ref{t1}). It is also worthwhile to consider the connection
between these configurations (with internal structures) and the dynamics of
the gauged $CP(2)$ vortices in the presence of terms representing the effect
of magnetic impurities (as studied for the Maxwell-Higgs case in the Ref. 
\cite{sk}), both scenarios presenting an explicit dependence of the Lagrange
density on the radial coordinate $r$. These issues are presently under
consideration, and we hope positive results for an incoming contribution.

\begin{acknowledgments}
This study was financed in part by the Coordena\c{c}\~ao de Aperfei\c{c}%
oamento de Pessoal de N\'{\i}vel Superior - Brasil (CAPES) - Finance Code
001. It was also supported by Brazilian Government via the Conselho Nacional
de Desenvolvimento Cient\'{\i}fico e Tecnol\'{o}gico (CNPq) and the Funda%
\c{c}\~{a}o de Amparo \`{a} Pesquisa e ao Desenvolvimento Cient\'{\i}fico e
Tecnol\'{o}gico do Maranh\~{a}o (FAPEMA). In particular, JA thanks the full
support from CAPES (postgraduate scholarship), RC acknowledges the support
from the grants CNPq/306385/2015-5 and FAPEMA/Universal-01131/17, and EH thanks the support from the grants CNPq/307545/2016-4 and CNPq/449855/2014-7. EH also acknowledges the School of Mathematics, Statistics and Actuarial Science of the University of Kent (Canterbury, United Kingdom) for the kind hospitality during the realization of part of this work.
\end{acknowledgments}

\end{document}